*Research Article*

# Regime-Switching Temperature Dynamics Model for Weather Derivatives

**Samuel Asante Gyamerah** ,[1] **Philip Ngare**,[2] **and Dennis Ikpe**[3]

[1]*Pan African University, Institute of Basic Sciences, Technology, and Innovation, Kenya*
[2]*University of Nairobi, Kenya*
[3]*University of South Africa, South Africa*

Correspondence should be addressed to Samuel Asante Gyamerah; saasgyam@gmail.com

Weather is a key production factor in agricultural crop production and at the same time the most significant and least controllable source of peril in agriculture. These effects of weather on agricultural crop production have triggered a widespread support for weather derivatives as a means of mitigating the risk associated with climate change on agriculture. However, these products are faced with basis risk as a result of poor design and modelling of the underlying weather variable (temperature). In order to circumvent these problems, a novel time-varying mean-reversion Lévy regime-switching model is used to model the dynamics of the deseasonalized temperature dynamics. Using plots and test statistics, it is observed that the residuals of the deseasonalized temperature data are not normally distributed. To model the nonnormality in the residuals, we propose using the hyperbolic distribution to capture the semiheavy tails and skewness in the empirical distributions of the residuals for the shifted regime. The proposed regime-switching model has a mean-reverting heteroskedastic process in the base regime and a Lévy process in the shifted regime. By using the Expectation-Maximization algorithm, the parameters of the proposed model are estimated. The proposed model is flexible as it modelled the deseasonalized temperature data accurately.

## 1. Introduction

From tilling of the farmland to selling of the output of the crop yield, farmers around the world make countless decisions that affect their performance. Yet, there is one very important factor that they cannot control, climate. The world's climate keeps on changing and this change will persist at rates that are projected to be out of the ordinary for some centuries [1]. Africa is no exception of these extreme climate changes across the world. Extreme climate events cause strain on food security, water resources, and human health in Africa. Ordinarily, it is the cause of limited economic growth and obstructs poverty reduction efforts for most countries in Africa [2]. With agriculture being the major contributing factor of the gross domestic product (GDP) growth of most countries in Africa [3] and climate conditions having extensive and causal correlation with the production variables [4], there should be an effective management technique to hedge against agricultural production risk. Most agricultural producers have encountered crop failures because of extreme weather conditions due to changes in climate. As a result, most farmers in Africa have developed their own traditional ways to improve the effect of extreme weather changes.

As an institutional response to weather changes, the Chicago Mercantile Exchange (CME) introduced the weather derivative (WD). WD has been in existence in most developed countries (Canada, Europe, USA, and Japan). However, most farmers in Africa have rarely heard about this effective hedging tool. WD, if introduced in Africa, will be more viable, reliable, and efficient to the agricultural industry and can hedge against the increasing weather changes that affect agriculture since it is devoid of factors like loss adjustments, moral hazards, adverse selections, high premiums, and complex information requirements. To avoid basis risk associated with WD, there should be an efficient model for the underlying weather variable used in pricing WD. The weather variable considered in this study is temperature. Temperature controls and influences other elements of weather like clouds, humidity, air pressure, and precipitation that affect crops during and after crop production (https://en.wikibooks.org/wiki/Basic Geography/Climate/Climate Elements, accessed on 26/01/2018.)

In the last decade, there has been empirical literatures on modelling the dynamics of temperature. Dischel [5] was the first to propose a continuous stochastic model for temperature. He modelled temperature as a mean-reverting process by adapting directly the Hull-White model. The noise process in his model was driven by two Wiener processes corresponding to the distribution of the temperature and the distribution of the changes in temperature. Thereafter, McIntyre and Doherty [6] proposed a mean-reverting SDE with a constant volatility daily average temperature at Heathrow airport in the United Kingdom (UK). Dornier and Queruel [7] disagreed with the direct use of the Hull-White model adopted by Dischel. They rather used a conventional Autoregressive Moving Average model rather than the AR(1) model proposed by Dischel. By replacing Brownian motion with fractional Brownian motion, Brody et al. [8] modelled the evolution of temperature that allowed the integration of a long memory effect. Other researchers [9–11] used different kinds of mean-reverting OU model driven by Brownian motion. In contrast to other researchers using Brownian motion to capture the residuals, Benth and Šaltytė-Benth [12] proposed an OU model that incorporate seasonal volatility and mean. In the model of Benth and Šaltytė-Benth, the residuals were driven by the generalized hyperbolic Lévy process rather than Brownian motion. The process they employed was an adjustable class of Lévy process that captured the skewness and semiheavy tails properties of the residuals.

Clearly, it can be observed that most of these researchers assumed no changes in state of the dynamics of temperature and hence modelled temperature dynamics as a single regime. The above methods of modelling temperature may lead to intractable pricing techniques for temperature derivatives. As noted by Brockett et al. [13], temperature time-series data shows sudden changes due to artificial and natural factors. By employing regime-switching models, the researcher can capture such sudden and discrete shifts in the temperature dynamics. Regime-switching models do capture most of the stylized facts of temperature accurately more than the single stochastic differential equation model, hence the need for different stochastic model for each switching state.

With a mean-reverting process as their base regime and a Brownian motion with mean different from zero as their shifted regime, Elias et al. [14] presented a constant volatility two-state MRS model for temperature dynamics at the city of Toronto, Canada. The model of Elias et al. (from hence we will call the model developed by Elias et al. as Elias' Model) failed to capture the fact that volatility of temperature varies with varying temperature as it goes through discrete changes between the states of the regime process. Evarest et al. [15] improved on Elias' model by capturing the fact that volatility of temperature varies as temperature goes through discrete changes between the states of the regime. They priced weather derivatives contracts based on the daily temperature dynamics. They used their model to calculate the future contract of HDD, CDD, and CAT indices. The introduction of the local volatility in the base regime helped in capturing well the dynamics of the underlying process. This led to a better pricing process as compared to Elias' model. However, they failed to capture the extreme and fat tail characteristics of temperature data in their model. In his seminal thesis, Cui [16] modelled and priced temperature derivative. He modelled the dynamics of temperature by a standard mean-reverting Ornstein-Uhlenbeck process with a general Lévy process as the driving noise. He extended his model by proposing a continuous-time autoregressive (CAR) model driven by a general Lévy process which he calibrated to the Canadian data. The two models he proposed were used in deriving futures price on HDD, CDD, and CAT. He later developed a two-state MRS model with a "normal" regime and a "jump" regime. The "normal" regime depended on a standard OU process. For the "jump" regime, he used different noise process (Brownian motion with more extreme drift and volatility) to drive the abnormal positive or negative "jumps" in the temperature dynamics. However, he failed to capture the changes in volatility of temperature during the MRS model but rather assumed a constant volatility in both regimes.

Several models have been formulated over time to capture the stylized facts of temperature; however these models proposed in literature have failed to capture well the stylized features of temperature, thus affecting the pricing models of WD. Inaccurate representation of the dynamics of temperature affects the pricing of WD. WD also relies on accurate extensive long-term time-series data [17]. However, there is lack of accessible, accurate, complete, and usable weather data in most African countries. Calibrating the MRS model is not trivial because the regimes are not clearly observable but latent. To outwit these problems, we use Expectation-Maximization (EM) algorithm to estimate the parameters in the model.

From the above literatures presented, Brownian motion has been replaced with a fractional Brownian motion and subsequently by a generalized hyperbolic Lévy processes. Nevertheless, it will be interesting to explore both Brownian motion and Lévy processes in a MRS model that incorporates "normal" temperatures and "extremes" in temperature. The contribution of this paper is twofold; firstly we developed a mathematically tractable temperature dynamics model for the African farmer by using regime-switching model and secondly we showed that Gaussian distribution cannot capture the dynamics of real-life temperature. To the best of our knowledge, the two-state regime-switching model developed is the first kind of model that can be used to price futures and options on futures.

## 2. Daily Temperature Dynamics

The most widely used temperature indices in most industries (energy consumers, energy industry, travel, transportation, agriculture, government, retailing, and construction) are the cumulative average temperature (CAT), cooling degree days (CDD), and heating degree days (HDD). Nevertheless, in this research, we use the CAT and growing degree days (GDD) since they are the dominant indices that affect agriculture in Africa [18, 19]. GDD is the measure of the suitability for a crop to grow in relation to the standard temperature.

*Definition 1.* For a given single temperature weather station, let $T^{max}(t)$ and $T^{min}(t)$ represent the daily maximum and minimum temperature (the temperatures used in this research are measured in degrees Celsius) recorded at day $t$, respectively. We define the daily average temperature at day $t$ as

$$T(t) \equiv \frac{T^{max}(t) + T^{min}(t)}{2}. \quad (1)$$

*Definition 2.* Assume the daily average temperature (DAT) $T(t)$ at time $t \geq 0$, and then the $CAT_t$ and $GDD_t$ generated at a specific location over a specific measurement period $[\tau_1, \tau_2]$ are defined as

$$CAT(\tau_1, \tau_2) := \sum_{t=\tau_1}^{\tau_2} T(t) \quad (2)$$

$$GDD(\tau_1, \tau_2) := \sum_{t=\tau_1}^{\tau_2} \max\{T(t) - T^{optimal}, 0\} \quad (3)$$

### 2.1. Stylized Facts of Temperature.
Temperature has clear characteristics which differs largely from commodities and other financial assets. The most palpable characteristics of temperature are the following.

*(i) Seasonality Feature.* Temperature exhibits annual (365 days) seasonal movements. The DAT $T(t)$ at time $t \geq 0$ is defined as the sum of the deseasonalized temperature $\widetilde{T}(t)$ and deterministic seasonal component $S_d(t)$ given as

$$T(t) = \widetilde{T}(t) + S_d(t) \quad (4)$$

To model the variations of temperature without the deterministic seasonality, the seasonal component in (4) will be removed to obtain the deseasonalized temperature $\widetilde{T}(t)$. The deterministic seasonal model at time $t$, $S_d(t)$, is defined as

$$S_d(t) = A_0 + A_1 t + A_2 \sin\left(\frac{2\pi}{365}(t - \varphi)\right) \quad (5)$$

where $A_0$ and $A_1$ represent the constant and coefficient in the linear seasonal trend of the raw data, respectively, $A_2$ captures the amplitude of the variation, and $\varphi$ is the phase angle.

*(ii) Mean-Reverting Feature.* It is practically impossible for daily temperature to deviate from the mean temperature over a long period. Daily temperature reverts toward the mean, a feature that is common to other commodities. As observed by Alaton et al. [9], long-term changes may be as a result different factors which includes but are not limited to global warming, green-house effects, and urbanization.

*(iii) Extreme Feature.* Temperature data have extremal data points. These extremal data points are "abnormal" movement caused by abrupt changes in temperature. In contrast to stocks which usually exhibit jumps in their price movements, daily temperature can show some signs of spikes which are normally short-lived and of very extreme size.

*(iv) Locality Feature.* Temperature has a strongly localized response in temperature modelling and as such requires caution in making generalization, hence the need for different models to capture these different characteristics at different locations.

*(v) Volatility.* In their two-state regime switching model formulation, Elias et al. [14] considered a constant volatility in either sate of his model. But this assumption might not be a reality since a shift in temperature residuals from one state to the other causes a change in the volatility from one state to the other. Extremal data points in temperature residues have greater volatility effects than is the case when there are no spikes or sudden increase in temperature. This is ascertained in the Engle test performed to check for heteroscedasticity in the temperature residuals (see Table 2). In model (9), the volatility is assumed to be dependent on the current deseasonalized temperature $\widetilde{T}(t)$. More precisely, the higher the deseasonalized temperature level, the larger the changes in the deseasonalized daily average temperature. Hence, in this study we will propose a model whose volatility differ with each regime and underlying process.

## 3. Markov Regime-Switching (MRS) Model

The Markov switching model developed by Hamilton [20] and Hamilton [21] inferred that the distribution of a variable is known, conditional on the occurrence of a specific regime/state. The switching process between the regimes is Markovian and is determined by an unobserved random variable. The underlying regimes, however, do not necessarily have to be Markovian but should be independent. The daily temperatures do change from day to day and these changes are not directly observable but latent. Therefore, statistical inference with regard to the likelihood of occurrence of each of the regimes at any time should be drawn.

MRS has been used effectively in modelling the behaviour of the stock market and spot price of electricity [22–25]. Chevallier and Goutte [26] used sixteen international stock markets to compare the performance of regime-switching Lévy models. Chevallier and Goutte [27] developed an estimation methodology that provided a better fit for electricity and $CO_2$ market prices by using mean-reverting Lévy jump processes.

In temperature modelling, it is typical to assume that there are different regimes that can capture distinct principal weather condition or the localized weather behaviour. In our study, the daily temperature is assumed to be latent with two possible regimes, either in the base regime ("normal or mean-reverting regime" $S_t = 1$) or in the shifted regime ("extreme" regime $S_t = 2$). Suppose that each regime in the regime-switching model undergoes discrete shifts between the regimes $S_t$ of the process, and then $S_t$ follows a first-order Markov process with the transition matrix:

$$\mathbf{P} = \begin{bmatrix} p_{11} & p_{12} \\ p_{21} & p_{22} \end{bmatrix} = \begin{bmatrix} p_{11} & 1 - p_{11} \\ 1 - p_{22} & p_{22} \end{bmatrix} \quad (6)$$

The transition probabilities of our temperature process $p_{ij}$ in (6) is given as

$$p_{ij} = \mathbf{P}(S_t = j \mid S_{t-1} = i) \quad \forall i, j = 1, 2 \quad (7)$$

$$0 \leq p_{ij} \leq 1$$

$$\text{and } \sum_{j=1}^{2} p_{ij} = 1 \quad (8)$$

Due to the Markov property of the states at any given time $t$, the future state of the underlying process (temperature) $S_{t+1}$ is independent of the past state $S_{t-1}$ of the underlying process given the present state $S_t$ of the underlying process.

*3.1. Modelling Daily Temperature Dynamics.* To efficiently model the dynamics of temperature, it is assumed that the deseasonalized temperature is either under base regime or shifted regime and each regime is independent and parallel to the other regimes. The deseasonalized temperature $\widetilde{T}(t)$ is assumed to be driven by two sources of randomness: a Markov process and Lévy process. We assume a constant mean-reversion rate in the base regime. Based on the stylized facts of temperature, a regime-switching stochastic model that describes the dynamics of temperature is formulated. This model can be used to price weather derivatives. The base regime model is assumed to follow a mean-reverting stochastic process with a time-varying volatility. The residuals of the base regime are assumed to be generated by a Brownian process.

To effectively capture the nonnormality of the temperature residuals (see Figures 3, 5, 6, and 7 and Table 5), the residuals of the shifted regime are captured by a Lévy process. By comparing the generalized hyperbolic distribution to its subclasses (normal-inverse Gaussian, Hyperbolic, and Variance-Gamma), we were able to find the best distribution that can model the asymmetry and heavy tails of the residuals data. As our first regime-switching model, we call it time-varying mean-reversion Lévy (TML) regime-switching model. This proposed model is distinctly appropriate to capture the dynamics of temperature. In sequel, the propose TML model for the deseasonalized temperature dynamics is given as

$$\widetilde{T}(t)$$

$$= \begin{cases} \widetilde{T}_t^M : d\widetilde{T}_t^M = \kappa \widetilde{T}_t^M dt + \sigma^M \widetilde{T}_t^M dW_t, & \text{with probability } p_1 \\ \widetilde{T}_t^L : d\widetilde{T}_t^L = \mu^L dt + \sigma^L dL_t, & \text{with probability } p_2 \end{cases} \quad (9)$$

where $\sigma^M \widetilde{T}_t^M$ is deseasonalized daily volatility of the base through time and $\sigma^L$ is the volatility of the shifted regimes and $\kappa$ is the mean-reversion rate of the deseasonalized temperature in the base regime which reverses the deseasonalized temperature to the long-term equilibrium level after the deseasonalized temperature has drifted from this equilibrium. $W_t \sim N(0, t)$ is the standard Brownian motion. $L_t$ is a Lévy process which is càdlàg, adapted, real-valued general Lévy process with independent, stationary increments and stochastically continuous, and $\widetilde{T}_t$ is the deseasonalized temperature at time $t$.

**Proposition 3.** *If the deseasonalized daily average temperature $\widetilde{T}(t)$ follows model (9), then the explicit solution is given by*

$$\widetilde{T}(t)$$

$$= \begin{cases} \widetilde{T}^M(t) : \widetilde{T}^M(t) = \widetilde{T}(t-1)e^{\kappa t} + \int_{t-1}^{t} \sigma \widetilde{T}(s) e^{\kappa(t-s)} dW(s) \\ \widetilde{T}^L(t) : \widetilde{T}^L(t) = \widetilde{T}(t-1) + \mu^L t + \int_{t-1}^{t} \sigma^L dL(s) \end{cases} \quad (10)$$

*Proof.* Determining the stochastic integral of the base regime process demands a variation of parameters approach to spell out a new function $f[T(t), t] = \widetilde{T}^W(t) e^{-\kappa t}$. By Itô's lemma, the derivative of the new function can be found.

$$df(T(t), t) = -\kappa e^{-\kappa t} \widetilde{T}^M(t) d(t) + e^{-\kappa t} d\widetilde{T}^M(t)$$

$$= -\kappa e^{-\kappa t} \widetilde{T}^M(t) d(t)$$

$$+ e^{-\kappa t} \left[ \kappa \widetilde{T}^M(t) dt + \sigma_1 \widetilde{T}^M(t) dW(t) \right]$$

$$= \sigma_1 e^{-\kappa t} \widetilde{T}(t) dW(t) \quad (11)$$

$$f(T(t), t) = \widetilde{T}^M(t) e^{-\kappa t}$$

$$= T(t-1) + \int_{t-1}^{t} \sigma_1 T(s) e^{-\kappa s} dW(s)$$

$$\widetilde{T}^M(t) = T(t-1) e^{\kappa t} + \int_{t-1}^{t} \sigma_1 T(s) e^{\kappa(t-s)} dW(s)$$

For the shifted regime

$$d\widetilde{T}^L(t) = \mu^L dt + \sigma_2 dL(t)$$

$$\int_{t-1}^{t} d\widetilde{T}^L(s) = \int_{t-1}^{t} \mu^L ds + \int_{0}^{t} \sigma_2 dL(s) \quad (12)$$

$$\widetilde{T}(t) = \widetilde{T}(t-1) + \mu^L t + \int_{t-1}^{t} \sigma_2 dL(s)$$

□

## 4. Analysis of Temperature Data

The daily maximum and minimum surface temperature data were taken from the weather measurement stations at Bole and Tamale. Bole and Tamale are located in the

Table 1: Descriptive statistics for daily average temperature.

|  | Mean | Median | Mode | std | Min | Max | Skewness | Kurtosis | Hurst Exponent | $\chi^2$ | P value |
| --- | --- | --- | --- | --- | --- | --- | --- | --- | --- | --- | --- |
| Bole | 27.20 | 27.00 | 26.5 | 2.07 | 21.40 | 34.20 | 0.41 | 2.77 | 0.8212 | 615.01 | 0 |
| Tamale | 28.69 | 28.40 | 27.8 | 2.47 | 21.40 | 35.60 | 0.31 | 2.34 | 0.7643 | 567.85 | 0 |

Table 2: Engle test for residual heteroscedasticity of Bole and Tamale.

|  | Test statistics | p-value |
| --- | --- | --- |
| Bole | 1175.2 | 0 |
| Tamale | 1282.9 | 0 |

Northern region (the hottest region in Ghana) of Ghana. Bole and Tamale are the district capital of Bole and Tamale, respectively. In Ghana, the main source of weather data is from the Ghana Meteorological Agency. The sample period expands from 01/01/1987 to 31/08/2012 and consists of a total of 9375 observations. The average of the daily maximum and minimum temperature is calculated according to Definition 1. The raw data is checked for missing data to avoid gaps in the historical data. Depending on the size of the missing data (the proportion of the missing data should not be more than 10%), the missing data is filled using the method of combined average. The combined average is calculated using two distinct averages: the average of 7 days (d) after and before the missing day,

$$T_{day}(t) = \frac{\sum_{d=1}^{7} T(t-d) + \sum_{d=1}^{7} T(t+d)}{14}, \quad (13)$$

and the average of that missing day across previous $N$ years,

$$T_{year}(t) = \frac{1}{N} \sum_{y=1}^{N} T_y(t) \quad (14)$$

The missing values in the dataset are filled by averaging the calculated value in (13) and (14).

In Table 1, the descriptive statistics for the daily average temperature of the two measurement stations (Bole and Tamale) are presented. The values of the median, mean, maximum, and minimum temperature for both towns are consistent and this can be attributed to the fact that the geographical locations of these two measurement stations are not distant apart. The amount of variation (std) is relatively small but vary between two measurement stations. With a skewness value of 0.41 and 0.31 for Bole and Tamale, respectively, the empirical distribution of these two towns are asymmetrical. With a negative excess kurtosis for both towns, it can be explained that the distribution of the DAT data is more outlier-prone than the normal distribution. We present the values $\chi^2$-statistics of Pearson's criteria of goodness-of-fit with its P values of the DAT time-series data (see Table 1). From the values of the $\chi^2$ goodness-of-fit test and at a 1% $\alpha$-level of significance, the null hypothesis (DAT data is normally distributed) can be rejected. With a Hurst exponent (H) greater than 0.5 for the two towns, there is a strong trend in the DAT data. However, the trend in Bole DAT data is more predictable than that of Tamale.

*4.1. Seasonal Component.* Generally, temperature follows a seasonal pattern. These seasonal patterns can be decomposed into a seasonal trend and linear trend depending on the region where the data was taken and the number of years of temperature data used. The seasonal component in the DAT time-series data is captured in model (5). In order to calibrate the DAT time-series data to the proposed model (9), the DAT time series will be deseasonalized. Figure 2 shows the deseasonalized data of the daily average temperature for Bole and Tamale.

From Figure 1, there is a strong seasonality in the DAT time-series data. Temperature exhibits a seasonal trend in Africa; higher temperatures during the dry seasons; and lower temperatures during the rainy seasons. The seasonal trend is captured in model (5). The seasonal component of model (5) is given as $A_2 \sin((2\pi/360)(t - \varphi))$. The DAT data is calibrated to the model and, using the least square method of estimation, the parameters are estimated.

Even though the linear trend in the data is weak, a close observation (see Figure 1) shows an increasing trend in the DAT time-series data. The linear trend in our data may be as a result of the longer period of years taken. Also, as observed by Alaton et al. [9], this increasing trend of DAT over some decades at this location is as a result of many factors which include but are not limited to global warming, green-house effect, and urbanization. The linear component of model (5) is given as $A_0 + A_1 t$.

*4.2. Residuals.* Figure 4 shows the squared residual plot of the deseasonalized daily average temperature data of Bole and Tamale. There is evidence of continuous variation in the variance of noise, an indication of seasonal heteroscedasticity. This shows that volatility of the deseasonalized DAT residuals is not constant as assumed by Elias et al. [14]. To further validate this result, we use the Engle test of residual heteroscedasticity to test for conditional heteroscedasticity. From Table 2, we can reject the null hypothesis of no conditional heteroscedasticity and conclude that there are significant heteroscedasticity effects in the residual series of both towns. With a p value of 0 and at a 1% $\alpha$-level of significance for both towns, there is a strong evidence to reject the null hypothesis of no conditional heteroscedasticity effects.

*4.2.1. Normality Test of Temperature Data Residuals.* In this section, statistical and graphical methods are applied to the DAT residuals time-series data to test for the normality of the residuals.

Most conventional literatures [9, 14, 15] assumed that the residuals are independent and identically and normally distributed. It should however be noted that inaccurate choice of the distribution of this residuals can cause model error and

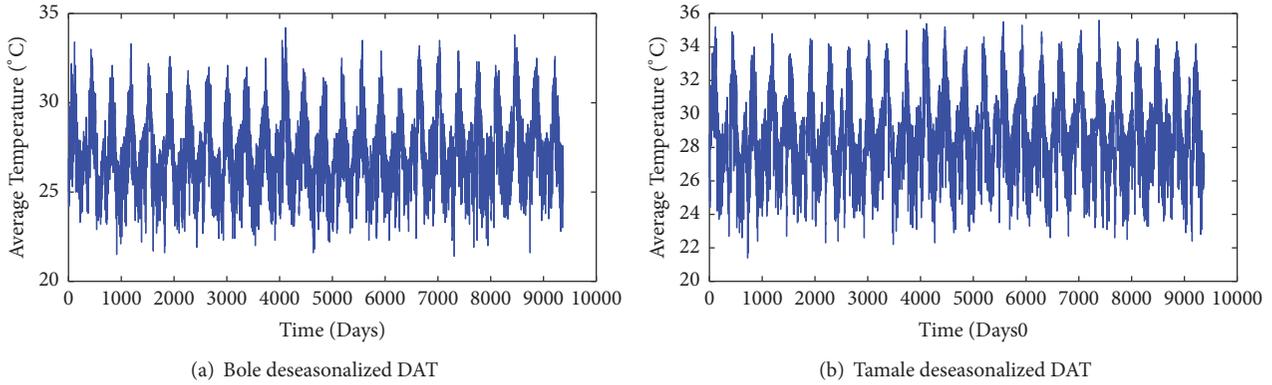

Figure 1: Historical average temperature against the day of observation from 01/01/1987 to 31/08/2012, exhibiting seasonal cycles.

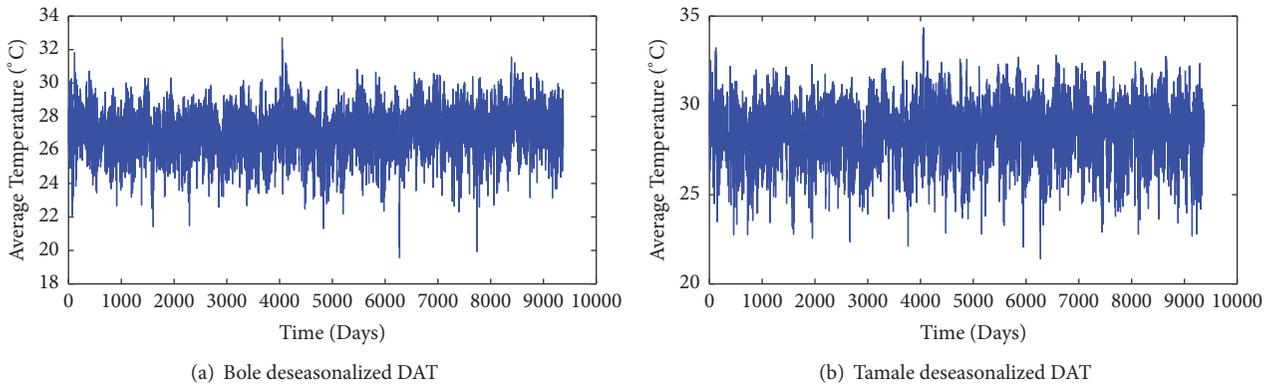

Figure 2: Deseasonalized daily average temperature from 01/01/1987 to 31/08/2012.

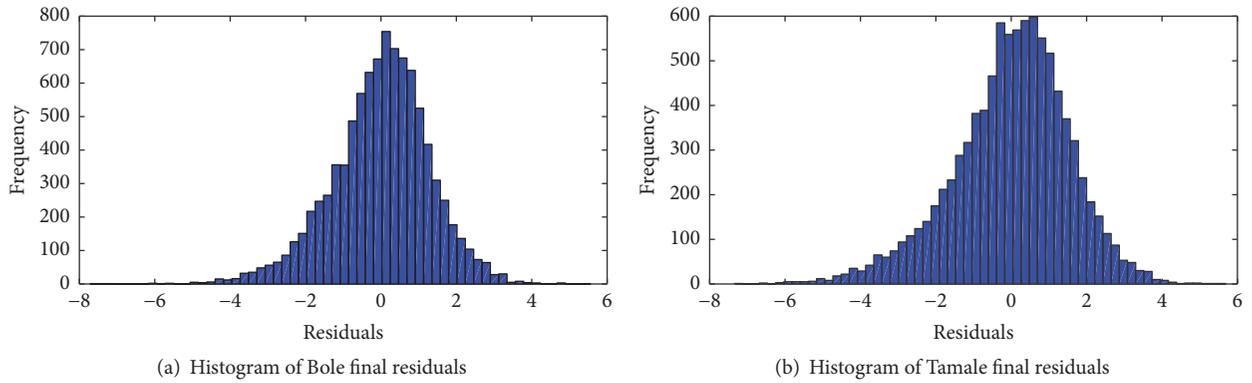

Figure 3: Histogram of final residuals in Bole and Tamale.

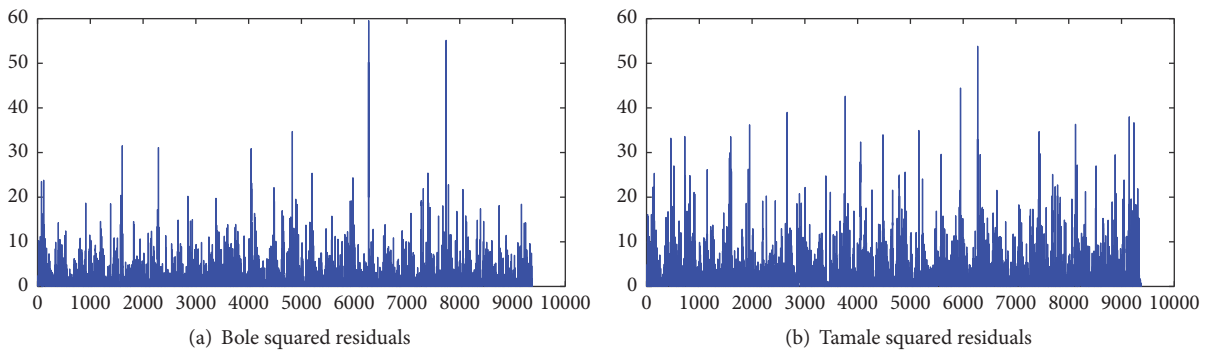

Figure 4: Squared final residuals of Bole and Tamale.

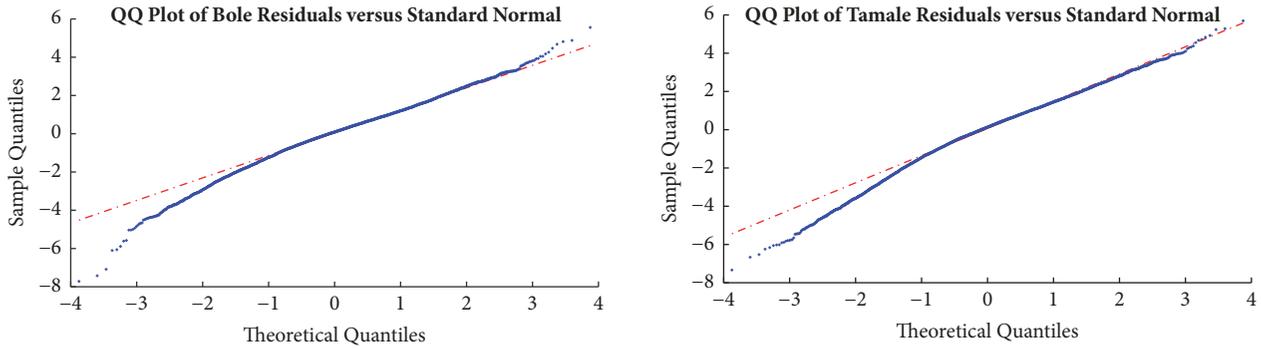

Figure 5: Normal distributed QQ-plot for Bole and Tamale residuals.

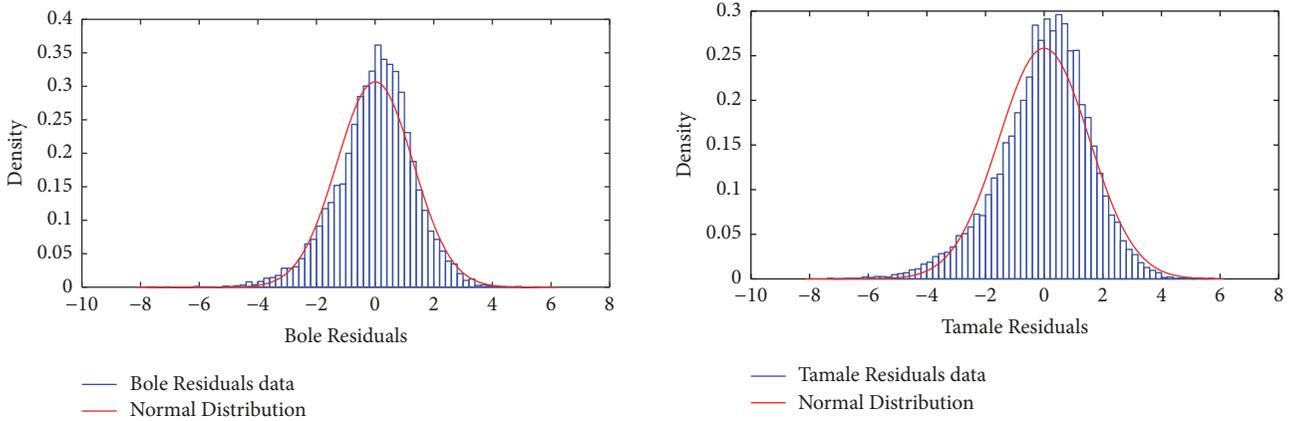

Figure 6: Normal fit plot of Bole and Tamale residuals.

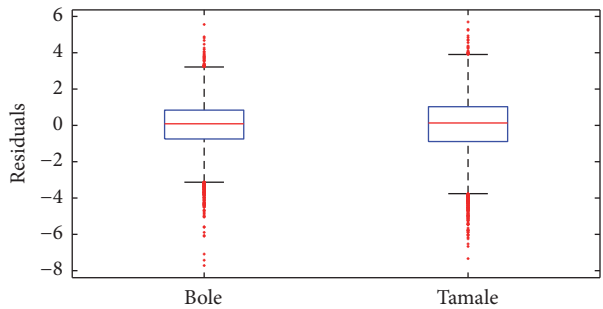

Figure 7: Box-plot of final residuals.

mispricing when pricing weather derivatives. To model the residuals of the DAT data, it is important to test the residuals for normality using different goodness-of-fit test.

Table 3 shows the normal distribution $\chi^2$-statistics of the residuals. The values of $\chi^2$ for Bole and Tamale are significant at 99% confidence level implying that the residuals of Bole and Tamale are not normally distributed. The descriptive statistics of the residuals of Bole and Tamale are presented in Table 4. From the Hurst exponent in Table 4, the residuals of Bole and Tamale DAT have strong trend and are more predictable. The skewness and kurtosis in Table 4 further show that the residuals are not normally distributed. These results are congruous to the Q-Q plot for normal distribution (Figure 5) and the normal fit plot (Figure 6). Figure 7 shows that there are cases of outliers in the residuals of the daily average temperature of Bole and Tamale.

Table 3: $\chi^2$ statistics of final residuals of Bole and Tamale.

| | $\chi^2$ | P value |
| --- | --- | --- |
| Bole | 230.3531 | 8.9851 ×10$^{-48}$ |
| Tamale | 419.0531 | 2.2355 ×10$^{-87}$ |

Additionally, using the Jacque-Bera (JB) goodness-of-fit test and the Anderson-Darling (AD) test for normality (see Table 5), we can reject the null hypothesis that the residuals

Table 4: Descriptive statistics of final residuals of Bole and Tamale.

|  | Mean | Median | std | Min | Max | Skewness | Kurtosis | Hurst Exponent |
|---|---|---|---|---|---|---|---|---|
| Bole | $1.0873 \times 10^{-06}$ | 0.0879 | 1.2995 | -7.7166 | 5.5563 | -0.4181 | 3.9910 | 0.7718 |
| Tamale | $3.6874 \times 10^{-06}$ | 0.1328 | 1.5431 | -7.3353 | 5.6889 | -0.4947 | 3.6699 | 0.7029 |

Table 5: Goodness-of-fit test for residuals.

(a) Jacque-Bera Test

|  | Bole | Tamale |
|---|---|---|
| Test Statistics | 656.7203 | 557.6335 |
| P value | $\leq 0.001$ | $\leq 0.001$ |

(b) Anderson-Darling test

|  | Bole | Tamale |
|---|---|---|
| Test Statistics | 26.0323 | 36.0366 |
| P value | $\leq 0.0005$ | $\leq 0.0005$ |

are normally distributed at a 99% confidence level. From the test statistics (Table 5) and graphical representations (Figures 5–7), there is enough evidence to state that the deseasonalized DAT residuals of Bole and Tamale do not follow the normal distribution. Thus, it can be concluded that it is not efficient to model the random noise with a Gaussian process.

Due to the inability of the normal distribution to capture well the residuals of the deseasonalized DAT data, we propose using the Lévy process to model the residuals of the shifted regime. We use the generalized hyperbolic (GH) distribution and its subclasses (normal-inverse Gaussian (NIG), hyperbolic (HYP), and the variance-gamma (VG)) to capture the skewness and semiheavy tails in the residuals data. We test for the best fit for our residuals data using the above named distributions.

### 4.3. Generalized Hyperbolic (GH) Distribution.
We model the residuals $\epsilon_t$ of the shifted regime by the generalized hyperbolic (GH) distribution and the subclasses which are relevant for applications.

*Definition 4.* The one-dimensional generalized hyperbolic (GH) distribution was introduced by Barndorff-Nielsen [28] and its probability density function is defined as

$$f_{GH}(x; \nu, \alpha, \beta, \mu, \delta) = \xi(\nu, \alpha, \beta, \delta) \left(\delta^2 + (x - \mu)^2\right)^{(\nu/2 - 1/4)} e^{\beta(x-\mu)} \\ \times K_{\nu-1/2}\left(\alpha \sqrt{\delta^2 - (x-\mu)^2}\right) \quad (15)$$

where $\xi(\nu, \alpha, \beta, \delta) = (\alpha^2 - \beta^2)^{\nu/2} / \sqrt{2\pi}\alpha^{\nu-1/2}\delta^\nu K_\nu(\delta\sqrt{\alpha^2 - \beta^2})$, $K_\nu$ is the modified Bessel function of the third kind of order $\nu$, and $x \in \mathbb{R}$

Each parameter in $GH(\nu, \alpha, \beta, \mu, \delta))$ distribution has different effect on the shape of the distribution: $\nu \in \mathbb{R}$ determines the characterization of subclasses of the GH distribution, $\alpha > 0$ controls the steepness around the peak (the larger $\alpha$, the steeper the density), $\beta$ with $0 \leq |\beta| \leq \alpha$ is the asymmetry parameter ($\beta = 0$ gives a symmetric distribution around $\mu$ and the skewness of the density increases as $\mu$ increases), $\mu$ is the location of the distribution, and $\delta > 0$ is the scaling. A proper choice of these parameters can help in describing different shapes of the distribution.

Suppose a random variable $X$ follows a generalized hyperbolic distribution, and then the characteristics function (moment generating function (MGF) or the cumulant function) is given as

$$M_X(z) = \mathbb{E}\left[e^{zX}\right] \\ = e^{\mu z} \left(\frac{\alpha^2 - \beta^2}{\alpha^2 - (\beta+z)^2}\right)^{\nu/2} \frac{K_\nu\left(\delta\sqrt{\alpha^2 - (\beta+z)^2}\right)}{K_\nu\left(\delta\sqrt{\alpha^2 - \beta^2}\right)}, \quad (16) \\ |\beta + z| < \alpha$$

The subclasses of the GH distribution include the normal-inverse Gaussian (*NIG*) distribution, the hyperbolic (*HYP*) distribution, and the variance-gamma (*VG*) distribution.

*Definition 5.* For $\nu = -0.5$ in the GH distribution, we obtain the normal-inverse Gaussian (*NIG*). The probability density function (pdf) of a NIG distribution $NIG(\alpha, \beta, \mu, \delta)$ of a random variable $X$ is an infinitely divisible distribution which is given as

$$f_{NIG}(x; \alpha, \beta, \mu, \delta)) = \alpha\delta\pi^{-1} \\ \cdot \exp\left\{\delta\sqrt{\alpha^2 - \beta^2} + \beta(x - \mu)\right\} \frac{K_1\left(\alpha\sqrt{\delta^2 + (x-\mu)^2}\right)}{\sqrt{\delta^2 + (x-\mu)^2}} \quad (17)$$

where $0 \leq |\beta| \leq \alpha$, $0 \leq \delta$, and $x, \mu \in \mathbb{R}$. $k_\nu$ is the modified Bessel function.

The NIG distribution was introduced in finance literature in 1997 by Barndorff-Nielsen [29]. NIG distribution has a heavier tail than the normal distribution and can take different kinds of shapes. The MGF of NIG distribution is

$$M_{NIG}(z) = \exp\left\{\mu z + \delta\left(\sqrt{\alpha^2 - \beta^2} - \sqrt{\alpha^2 - (\beta+z)^2}\right)\right\}, \quad (18) \\ \forall |\beta + z| < \alpha$$

The NIG distribution have the following properties:

(1) Conditioned that $\beta = 0$, $\alpha \longrightarrow \infty$, and $\sigma/\alpha = \sigma^2$, the NIG distribution will approach the normal distribution $N(\mu, \sigma^2)$.

(2) Assuming that $X \sim NIG(\alpha, \beta, \mu, \delta)$, then, $\forall a \in \mathbb{R}^+$ and $\forall b \in \mathbb{R}$, we have that $Y = aX + b \sim NIG(\alpha/a, \beta/a, a\mu + b, a\delta)$

*Definition 6.* The Hyperbolic Distribution (HYP) is a subclass of the GH distribution when $\nu = 1$. Given a hyperbolic random variable $X$, the pdf is given as

$$f_{HYP}(x; \alpha, \beta, \mu, \delta)) = \frac{\sqrt{\alpha^2 - \beta^2}}{2\alpha\delta K_1\left(\delta\sqrt{\alpha^2 - \beta^2}\right)} \quad (19)$$
$$\cdot \exp\left(-\alpha\sqrt{\delta^2 + (x-\mu)^2} + \beta(x-\mu)\right)$$

where $x, \mu \in \mathbb{R}, 0 \leq \delta, |\beta| < \alpha$ and $K_1$ is t the Bessel function of the third kind with index 1

The hyperbolic distribution can also capture semiheavy tails. The MGF of the hyperbolic distribution is

$$M_{HYP}(z) = \frac{e^{\mu z}\sqrt{\alpha^2 - \beta^2}K_1\left(\delta\sqrt{\alpha^2 - (\beta+z)^2}\right)}{\sqrt{\alpha^2 - (\beta+z)^2}K_1\left(\delta\left(\sqrt{\alpha^2 - \beta^2}\right)\right)}, \quad (20)$$
$$\forall |\beta + z| < \alpha$$

The hyperbolic distribution is not closed under convolution.

*Definition 7.* By restricting $\delta = 0$ and $\nu > 0$, we have the Variance-Gamma (VG) distribution with pdf

$$f_{VG}(x; \nu, \alpha, \beta, \mu)) = \frac{\left(\sqrt{\alpha^2 - \beta^2}\right)^{2\nu}}{\sqrt{\pi}\Gamma(\nu)(2\alpha)^{\nu-1/2}}|x-\mu|^{\nu-1/2} \quad (21)$$
$$\cdot K_{\nu-1/2}(\alpha|x-\mu|)e^{\beta(x-\mu)},$$
$$\forall x \in \mathbb{R}$$

where $\mu \in \mathbb{R}, \nu > 0, \alpha > |\beta| \geq 0$, and $\Gamma$ denotes the gamma-function. $K_\nu(\cdot)$ is the modified Bessel function of the second kind.

The Variance-Gamma density was introduced by Madan et al. [30]. The tails of VG distribution decrease slowly than the normal distribution. The class of VG distributions is closed under convolution. The MGF of VG distribution is

$$M_{VG}(z) = e^{\mu z}\left(\frac{\sqrt{\alpha^2 - \beta^2}}{\sqrt{\alpha^2 - (\beta+z)^2}}\right)^{2\nu}, \quad (22)$$
$$\forall |\beta + z| < \alpha$$

*4.4. Parameter Estimation.* The parameters of the seasonality model and the TML model is estimated. The estimation of these parameters depends on the Bole and Tamale DAT data from 01/01/1987 to 31/08/2012.

*4.4.1. Fitting the Deterministic Seasonality Model to Data.* The deterministic seasonality process, model (5), can be transformed to

$$S_d(t) = a_0 + a_1(t) + a_2 \sin\left(\frac{2\pi}{365}t\right) + a_3 \cos\left(\frac{2\pi}{365}t\right) \quad (23)$$

To find the numerical values in model (5), the transformed deterministic seasonality process (model (23)) is fitted to the DAT data. The parameters are estimated using the least square methods.

$$A_0 = a_0$$
$$A_1 = a_1$$
$$A_2 = \sqrt{(a_2^2 + a_3^2)} \quad (24)$$
$$\varphi = \frac{365}{2\pi}\tan^{-1}\left(\frac{a_3}{a_2}\right)$$

*4.4.2. Estimation of the TML Parameters.* Estimating the parameters of TML model is not trivial. The parameters of the model are estimated using the Expectation-Maximization algorithm developed by Dempster et al. [31]. The vector of unknown parameters will be estimated by two steps' iterative algorithm: an expectation- (E-) step and a maximization- (M-) step.

*Discretization.* Even though temperature is a continuous process, its data is not recorded continuously but rather recorded in discrete time points. Therefore, estimating the parameters in a continuous time will be computationally costly. Before the parameters of the TML model are estimated, they will have to be calibrated to the deseasonalized DAT data and this is possible by transforming the model from its continuous form to a discrete form. The discretized form of model (9) for the base and shifted regimes is given as

$$\widetilde{T}(t)$$
$$= \begin{cases} \widetilde{T}_t^M : \widetilde{T}_t^M = \widetilde{T}_{t-1}^M(1+\kappa) + \sigma^M \widetilde{T}_{t-1}^M \epsilon_t^M, & \widetilde{T}^M \text{ is in regime 1} \\ \widetilde{T}_t^L : \widetilde{T}_t^L = \widetilde{T}_{t-1}^L + \mu^L + \sigma^L \epsilon_t^L, & \widetilde{T}^L \text{ is in regime 2} \end{cases} \quad (25)$$

where $\epsilon^M(t)$ and $\epsilon^L(t)$ are the Weiner residual and Lévy residual, respectively.

From (25), the vector of unknown parameters $\theta_1 = \{\beta^M, \sigma^M, p_1\}$ and $\theta_2 = \{\mu^L, \sigma^L, p_2\}$ for the base and shifted regimes, respectively, will be estimated.

*E-Step.* Assume the length of the DAT historical data is $N + 1$ and $t = 0, 1, 2, 3, \ldots, N$ where $t$ represents a specific time that the DAT is recorded and $\theta^n$ is the computed vector of parameters in the $n^{th}$ iteration. The conditional distribution of the regimes $S_t$ for time update values of $t = 0, 1, 2, 3, \ldots, N$ will be calculated. Suppose $\mathscr{F}_t^{\widetilde{T}(t)}$ is a vector of the past $t + 1$ historical data of the discretized model, and then $\mathscr{F}_t^{\widetilde{T}(t)} = \{\widetilde{T}(1), \widetilde{T}(2), \widetilde{T}(3), \ldots, \widetilde{T}(N)\}$.

(i) *Filtering.* Based on the Bayes rule, the filtered probability of the discretized model can be estimated as

$$\mathbb{P}\left(S_t = i \mid \mathcal{F}_t^{\tilde{T}(t)}; \Theta^{(n)}\right) = \frac{\mathbb{P}\left(S_t = i, \tilde{T}(t) \mid \mathcal{F}_{t-1}^{\tilde{T}(t)}; \Theta^{(n)}\right)}{f\left(\tilde{T}(t) \mid \mathcal{F}_{t-1}^{\tilde{T}(t)}; \Theta^{(n)}\right)}$$

$$= \frac{\mathbb{P}\left(S_t = i \mid \mathcal{F}_{t-1}^{\tilde{T}(t)}; \Theta^{(n)}\right) f\left(\tilde{T}(t) \mid S_t = i; \mathcal{F}_{t-1}^{\tilde{T}(t)}; \Theta^{(n)}\right)}{\sum_{i \in s} \mathbb{P}\left(S_t = i \mid \mathcal{F}_{t-1}^{\tilde{T}(t)}; \Theta^{(n)}\right) f\left(\tilde{T}(t) \mid S_t = i; \mathcal{F}_{t-1}^{\tilde{T}(t)}; \Theta^{(n)}\right)} \quad (26)$$

where $\Theta = \{\theta_1, \theta_2\}$ and $f(\tilde{T}(t) \mid S_t = i; \mathcal{F}_{t-1}^{\tilde{T}(t)})$ is the density of the underlying regime process $i$ at time $t$ conditional that the underlying process was in regime $i$. The conditional probability density function for the base and shifted regimes will be calculated from the CDF.

From (25), the drift and diffusion coefficient of the base regime are $(1+\beta)\tilde{T}_{t-1}^M$ and $\sigma^M \tilde{T}_{t-1}^M$, respectively. Similarly, the drift and diffusion coefficient of the shifted regimes are $\tilde{T}_{t-1}^L + \mu^L$ and $\sigma^L$, respectively.

(ii) *Smoothing*. For $t = N-1, N-2, \ldots, 1$ iterate

$$\mathbb{P}\left(S_t = i \mid \mathcal{F}_t^{\tilde{T}(t)}; \Theta^{(n)}\right)$$

$$= \sum_{i \in s} \frac{\mathbb{P}\left(S_t = i \mid \mathcal{F}_t^{\tilde{T}(t)}; \Theta^{(n)}\right) \mathbb{P}\left(S_{t+1} = j \mid \mathcal{F}_t^{\tilde{T}(T)}; \Theta^{(n)}\right) p_{ij}^{(n)}}{\mathbb{P}\left(S_{t+1} = i \mid \mathcal{F}_t^{\tilde{T}(t)}; \Theta^{(n)}\right)} \quad (27)$$

The probability density functions (pdf) of the base and shifted regimes based on their diffusion and drift coefficient is, respectively, given as

$$f\left(\tilde{T}_t \mid S_t = i; \mathcal{F}_{t-1}^{\tilde{T}(t)}; \widehat{\theta}_1^{(n)}\right)$$

$$= \frac{1}{\sigma_1^n \sqrt{2\pi} \tilde{T}_{t-1}} \exp\left[-\frac{\left(\tilde{T}_t - (1+\kappa^{(n)}) \tilde{T}_{t-1}\right)^2}{2\left(\sigma_1^{(n)}\right)^2 \tilde{T}_{t-1}^2}\right] \quad (28)$$

$$f\left(\tilde{T}_t \mid S_t = i; \mathcal{F}_{t-1}^{\tilde{T}(t)}; \widehat{\theta}_2^{(n)}\right)$$

$$= \frac{1}{\sigma_2^n \sqrt{2\pi}} \exp\left[-\frac{\left(\tilde{T}_t - \mu^{(n)} - \tilde{T}_{t-1}\right)^2}{2\left(\sigma_2^{(n)}\right)^2}\right] \quad (29)$$

*M-Step*. By maximizing the expected log-likelihood function $\Theta^{(n+1)} = \text{argmax}\, Q(\Theta \mid \Theta^{(n)})$, the maximum likelihood (ML) estimate $\Theta^{(n+1)}$ for the vector of unknown parameters will be calculated. Also, in this step the transition probabilities of the regime-switching model will be estimated.

The log-likelihood function from the conditional probability density function of both regimes is, respectively, given as

$$\log\left[L\left(\theta_1^{(n)}, \mathcal{F}_t^{\tilde{T}}, S_t\right)\right] = \sum_{t=2}^N \mathbb{P}\left(S_t = i \mid \mathcal{F}_t^{\tilde{T}(t)}; \widehat{\theta}_1^{(n)}\right)$$

$$\cdot \left[\log P_{1,i} - \log\left(\sigma_1 \sqrt{2\pi} \tilde{T}_{t-1}\right)\right. \quad (30)$$

$$\left. - \frac{1}{2\sigma_1^2 \tilde{T}_{t-1}^2}\left(\tilde{T}_t - (1+\kappa^n)\tilde{T}_{t-1}\right)^2\right]$$

$$\log\left[L\left(\theta_1^{(n)}, \mathcal{F}_t^{\tilde{T}}, S_t\right)\right] = \sum_{t=2}^N \mathbb{P}\left(S_t = i \mid \mathcal{F}_t^{\tilde{T}(t)}; \widehat{\theta}_2^{(n)}\right)$$

$$\cdot \left[\log P_{2,i} - \log\left(\sigma_2 \sqrt{2\pi}\right)\right. \quad (31)$$

$$\left. - \frac{1}{2\sigma_2^2}\left(\tilde{T}_t - \mu^{(n)} - \tilde{T}_{t-1}\right)^2\right]$$

By maximizing the log-likelihood function presented in (30), the vector of parameters $\theta_1^{(n+1)}$ can be estimated.

$$\widehat{\sigma}_1^{(n+1)} = \sqrt{\frac{\sum_{t=2}^N \left[\left(\mathbb{P}\left(S_t = 1 \mid \mathcal{F}_t^{\tilde{T}(t)}; \widehat{\theta}_1^{(n)}\right)\right) \tilde{T}_{t-1}^{-2}\left(\left(\tilde{T}_t - (1+\beta)\tilde{T}_{t-1}\right)^2\right)\right]}{\sum_{t=2}^N \mathbb{P}\left(S_t = 1 \mid \mathcal{F}_t^{\tilde{T}(t)}; \widehat{\theta}_1^{(n)}\right)}} \quad (32)$$

$$\widehat{\kappa}^{(n+1)} = \frac{\sum_{t=1}^N \left[\left(\mathbb{P}\left(S_t = 1 \mid \mathcal{F}_t^{\tilde{T}(t)}; \widehat{\theta}_1^{(n)}\right)\right) \tilde{T}_{t-1}^{-2}\left(\tilde{T}_{t-1}\left(\tilde{T}_t - \tilde{T}_{t-1}\right)\right)\right]}{\sum_{t=2}^N \left[\mathbb{P}\left(S_t = 1 \mid \mathcal{F}_t^{\tilde{T}(t)}; \widehat{\theta}_1^{(n)}\right)\right]} \quad (33)$$

*Proof.* The first derivative of (30) with respect to $\sigma_1$ is

$$\frac{\partial L}{\partial \sigma_1} = \frac{1}{\sigma_1^3} \sum_{t=2}^N \left[\left(\mathbb{P}\left(S_t = 1 \mid \mathcal{F}_t^{\tilde{T}(t)}; \widehat{\theta}_1^{(n)}\right)\right)\right.$$

$$\left. \cdot \left(\frac{\left(\tilde{T}_t - (1+\beta)\tilde{T}_{t-1}\right)^2 - \sigma_1^2 \tilde{T}_{t-1}^2}{\tilde{T}_{t-1}^2}\right)\right] \quad (34)$$

By maximizing (34),

$$\sum_{t=2}^{N} \left[ \left( \mathbb{P}\left(S_t = 1 \mid \mathcal{F}_t^{\tilde{T}(t)}; \widehat{\theta}_1^{(n)}\right) \right) \tilde{T}_{t-1}^{-2} \left( \left(\tilde{T}_t - (1+\beta)\tilde{T}_{t-1}\right)^2 - \sigma_1^2 \tilde{T}_{t-1}^2 \right) \right] = 0$$

$$\sum_{t=2}^{N} \left[ \left( \mathbb{P}\left(S_t = 1 \mid \mathcal{F}_t^{\tilde{T}(t)}; \widehat{\theta}_1^{(n)}\right) \right) \tilde{T}_{t-1}^{-2} \left( \left(\tilde{T}_t - (1+\beta)\tilde{T}_{t-1}\right)^2 \right) \right] = \sum_{t=2}^{N} \mathbb{P}\left(S_t = 1 \mid \mathcal{F}_t^{\tilde{T}(t)}; \widehat{\theta}_1^{(n)}\right) \sigma_1^2 \quad (35)$$

$$\widehat{\sigma}_1^{(n+1)} = \sqrt{\frac{\sum_{t=2}^{N} \left[ \left( \mathbb{P}\left(S_t = 1 \mid \mathcal{F}_t^{\tilde{T}(t)}; \widehat{\theta}_1^{(n)}\right) \right) \tilde{T}_{t-1}^{-2} \left( \left(\tilde{T}_t - (1+\beta)\tilde{T}_{t-1}\right)^2 \right) \right]}{\sum_{t=2}^{N} \mathbb{P}\left(S_t = 1 \mid \mathcal{F}_t^{\tilde{T}(t)}; \widehat{\theta}_1^{(n)}\right)}}$$

The derivative of (30) with respect to $\beta$ is

$$\frac{\partial L}{\partial \kappa} = \sum_{t=1}^{N} \left[ \left( \mathbb{P}\left(S_t = 1 \mid \mathcal{F}_t^{\tilde{T}(t)}; \widehat{\theta}_1^{(n)}\right) \right) \cdot \sigma_1^{-2} \tilde{T}_{t-1}^{-2} \left( -\tilde{T}_{t-1}^2 - \beta \tilde{T}_{t-1}^2 + \tilde{T}_t \tilde{T}_{t-1} \right) \right] \quad (36)$$

By maximizing (36),

$$\sum_{t=1}^{N} \left[ \left( \mathbb{P}\left(S_t = 1 \mid \mathcal{F}_t^{\tilde{T}(t)}; \widehat{\theta}_1^{(n)}\right) \right) \cdot \sigma_1^{-2} \tilde{T}_{t-1}^{-2} \left( -\tilde{T}_{t-1}^2 - \beta \tilde{T}_{t-1}^2 + \tilde{T}_t \tilde{T}_{t-1} \right) \right] = 0$$

$$\kappa \sum_{t=2}^{N} \left[ \mathbb{P}\left(S_t = 1 \mid \mathcal{F}_t^{\tilde{T}(t)}; \widehat{\theta}_1^{(n)}\right) \right]$$

$$= \sum_{t=1}^{N} \left[ \left( \mathbb{P}\left(S_t = 1 \mid \mathcal{F}_t^{\tilde{T}(t)}; \widehat{\theta}_1^{(n)}\right) \right) \tilde{T}_{t-1}^{-2} \left( \tilde{T}_{t-1} \left( \tilde{T}_t - \tilde{T}_{t-1} \right) \right) \right] \quad (37)$$

$$\widehat{\kappa}^{(n+1)} = \frac{\sum_{t=1}^{N} \left[ \left( \mathbb{P}\left(S_t = 1 \mid \mathcal{F}_t^{\tilde{T}(t)}; \widehat{\theta}_1^{(n)}\right) \right) \tilde{T}_{t-1}^{-2} \left( \tilde{T}_{t-1} \left( \tilde{T}_t - \tilde{T}_{t-1} \right) \right) \right]}{\sum_{t=2}^{N} \left[ \mathbb{P}\left(S_t = 1 \mid \mathcal{F}_t^{\tilde{T}(t)}; \widehat{\theta}_1^{(n)}\right) \right]}$$

□

Also, the vector of unknowns of $\theta_2^{(n+1)}$ can be estimated by maximizing (31):

$$\widehat{\mu}^{(n+1)} = \frac{\sum_{t=2}^{N} \left[ \left( \mathbb{P}\left(S_t = 2 \mid \mathcal{F}_t^{\tilde{T}(t)}; \widehat{\theta}_2^{(n)}\right) \right) \left( \tilde{T}_t - \tilde{T}_{t-1} \right) \right]}{\sum_{t=2}^{N} \left[ \mathbb{P}\left(S_t = 2 \mid \mathcal{F}_t^{\tilde{T}(t)}; \widehat{\theta}_2^{(n)}\right) \right]} \quad (38)$$

$$\widehat{\sigma}_2^{(n+1)} = \sqrt{\frac{\sum_{t=2}^{N} \left[ \left( \mathbb{P}\left(S_t = 2 \mid \mathcal{F}_t^{\tilde{T}(t)}; \widehat{\theta}_2^{(n)}\right) \right) \left( \tilde{T}_t - \tilde{T}_{t-1} - \mu \right)^2 \right]}{\sum_{t=2}^{N} \left[ \mathbb{P}\left(S_t = 2 \mid \mathcal{F}_t^{\tilde{T}(t)}; \widehat{\theta}_2^{(n)}\right) \right]}} \quad (39)$$

*Proof.* By finding the derivative of (31) with respect to $\sigma_2$,

$$\frac{\partial L}{\partial \sigma_2} = \frac{1}{\sigma_2^3} \sum_{t=2}^{N} \left[ \left( \mathbb{P}\left(S_t = 2 \mid \mathcal{F}_t^{\tilde{T}(t)}; \widehat{\theta}_2^{(n)}\right) \right) \cdot \left( \left(\tilde{T}_t - \tilde{T}_{t-1} - \mu\right)^2 - \sigma_2^2 \right) \right] \quad (40)$$

Maximizing (40) gives

$$\sum_{t=2}^{N} \left[ \left( \mathbb{P}\left(S_t = 2 \mid \mathcal{F}_t^{\tilde{T}(t)}; \widehat{\theta}_2^{(n)}\right) \right) \left( \left(\tilde{T}_t - \tilde{T}_{t-1} - \mu\right)^2 - \sigma_2^2 \right) \right] = 0$$

$$\sum_{t=2}^{N} \left[ \left( \mathbb{P}\left(S_t = 2 \mid \mathcal{F}_t^{\tilde{T}(t)}; \widehat{\theta}_2^{(n)}\right) \right) \left( \left(\tilde{T}_t - \tilde{T}_{t-1} - \mu\right)^2 \right) \right]$$

$$= \sigma_2^2 \sum_{t=2}^{N} \left( \mathbb{P}\left(S_t = 2 \mid \mathcal{F}_t^{\tilde{T}(t)}; \widehat{\theta}_2^{(n)}\right) \right) \quad (41)$$

$$\widehat{\sigma}_2^{(n+1)} = \sqrt{\frac{\sum_{t=2}^{N} \left[ \left( \mathbb{P}\left(S_t = 2 \mid \mathcal{F}_t^{\tilde{T}(t)}; \widehat{\theta}_2^{(n)}\right) \right) \left( \tilde{T}_t - \tilde{T}_{t-1} - \mu \right)^2 \right]}{\sum_{t=2}^{N} \left[ \mathbb{P}\left(S_t = 2 \mid \mathcal{F}_t^{\tilde{T}(t)}; \widehat{\theta}_2^{(n)}\right) \right]}}$$

The derivative of (31) with respect to $\mu$ is

$$\frac{\partial L}{\partial \sigma_2} = \sum_{t=2}^{N} \left[ \left( \mathbb{P}\left(S_t = 2 \mid \mathcal{F}_t^{\tilde{T}(t)}; \widehat{\theta}_2^{(n)}\right) \right) \cdot \left( \sigma_2^{-2} \left( \tilde{T}_t - \tilde{T}_{t-1} - \mu \right)^2 \right) \right] \quad (42)$$

By maximizing (42)

$$\sum_{t=2}^{N} \left[ \left( \mathbb{P}\left(S_t = 2 \mid \mathcal{F}_t^{\tilde{T}(t)}; \widehat{\theta}_2^{(n)}\right) \right) \left( \left(\tilde{T}_t - \tilde{T}_{t-1}\right)^2 \right) \right]$$

$$= \mu \sum_{t=2}^{N} \left( \mathbb{P}\left(S_t = 2 \mid \mathcal{F}_t^{\tilde{T}(t)}; \widehat{\theta}_2^{(n)}\right) \right) \quad (43)$$

$$\widehat{\mu}^{(n+1)}$$

$$= \frac{\sum_{t=2}^{N} \left[ \left( \mathbb{P}\left(S_t = 2 \mid \mathcal{F}_t^{\tilde{T}(t)}; \widehat{\theta}_2^{(n)}\right) \right) \left( \tilde{T}_t - \tilde{T}_{t-1} \right) \right]}{\sum_{t=2}^{N} \left[ \mathbb{P}\left(S_t = 2 \mid \mathcal{F}_t^{\tilde{T}(t)}; \widehat{\theta}_2^{(n)}\right) \right]} \quad (44)$$

□

The transition probabilities are estimated by making use of the formula proposed by Kim [32]:

$$p_{ij}^{(n+1)} = \frac{\sum_{t=2}^{N} \mathbb{P}\left(S_t = j, S_{t-1} = i \mid \mathcal{F}_N^{\widetilde{T}(t)}; \Theta^{(n)}\right)}{\sum_{t=2}^{N} \mathbb{P}\left(S_t = i \mid \mathcal{F}_N^{\widetilde{T}(t)}; \Theta^{(n)}\right)}$$

$$= \frac{\sum_{t=2}^{N} \mathbb{P}\left(S_t = j \mid \mathcal{F}_N^{\widetilde{T}(t)}; \Theta^{(n)}\right)\left(p_{ij}^{(n)} \mathbb{P}\left(S_{t-1} = i\right) \mathcal{F}_{t-1}^{\widetilde{T}(t)}; \Theta^{(n)} / \mathbb{P}\left(S_t = j \mid \mathcal{F}_{t-1}^{\widetilde{T}(t)}; \Theta^{(n)}\right)\right)}{\sum_{t=2}^{N} \mathbb{P}\left(S_{t-1} = i \mid \mathcal{F}_N^{\widetilde{T}(t)}; \Theta^{(n)}\right)}$$

(45)

## 5. Discussion and Results

By inserting the estimated parameters into the transformed deterministic seasonality process (model (23)), the deterministic seasonal DAT is obtained for Bole and Tamale. The linear trend in the model is evidently very small. However, the linear trend of Bole DAT is more evident than that of Tamale DAT. Using the parameter estimates values in Table 6, seasonal sine graph is fitted to our Bole and Tamale DAT data (see Figure 8).

$$S_d(t) = 26.8194 + \left(2.3855 \times 10^{-05}\right) t$$
$$- 2.0234 \sin \frac{2\pi}{365} (t - 196.2153)$$
$$S_d(t) = 28.5058 + \left(3.7039 \times 10^{-05}\right) t$$
$$- 2.1026 \sin \frac{2\pi}{365} (t - 200.5695)$$

(46)

The estimated parameters of normal, HYP, GH, NIG, and VG distributions are presented in the Table 7.

To test for the goodness-of-fit of the distributions, two distance measures are used; the Kolmogorov-Smirnov (K-S) and the Anderson-Darling (A-D) statistic are used. The K-S test statistic and A-D test statistic are used to summarize the difference between the fitted cumulative density function (cdf) and the empirical cdf. Comparative to Kolmogorov-Smirnov, the A-D test statistics is more powerful because it incorporates integration over the entire range of data by paying more attention to the tail distances. The lower the value of K-S and A-D test statistics, the better the fit of that distribution.

Regardless of the test used, the distance between the fitted hyperbolic distribution and its empirical distribution is lower compared to the other distributions and their empirical. This affirms that the hyperbolic distribution does fit well to our Bole and Tamale random noise. A series of random numbers for GH, NIG, HYP, and VG are generated using the parameters estimated in Table 7. The Q-Q plots of the quantiles of the residuals versus the randomly generated quantiles of the GH, NIG, HYP, and VG distributions are plotted (see Figure 9). The straight line (in red) shows how the residual data would behave if it is perfectly distributed with the GH, NIG, HYP, and VG. From the illustrated figures (Figure 9), it is evident that the hyperbolic distribution fits plausibly well for our Bole and Tamale random components than the normal, GH, NIG, HYP, and VG distributions. This is consistent with the A-D and K-S goodness-of-fit test in Table 9.

The deseasonalized DAT $\widetilde{T}_t$ and the conditional probability of being in the extreme regime $\mathbb{P}(S_t = 2)$ for the historical deseasonalized DAT are shown in Figure 10. The deseasonalized DAT that are categorize as "extremes," that is, with $\mathbb{P}(S_t = 2) > 0.8$, are represented by red dots.

The estimates of the TML model for deseasonalized temperature is presented in Table 8. The speed of the mean-reversion is fairly low for both Bole and Tamale. However, the mean-reversion rate of Tamale is higher than that of Bole. The Markov probability $p_{11}$ of the deseasonalized DAT to stay in the "normal" regime of the TML model at Bole and Tamale is higher than the Markov probability $p_{22}$ of the deseasonalized DAT to stay in the "extreme" regime at Bole and Tamale. We can conclude that the "normal" regime of the TML model at both Bole and Tamale is relatively stable comparative to the "extreme" regime of the TML model. However, it is evident that there are instances that the temperatures of Bole and Tamale are at their extremes since the probability of staying in the "extreme" regime is significant even though it is small comparative to the probability of staying in the "normal" regime.

## 6. Conclusion

In this research, a novel time-varying mean-reversion Lévy regime-switching (TML) temperature dynamics model which captures the normal variations and extreme variations in temperature is developed to characterize the stochastic dynamics of temperature. The model includes a time-varying volatility and a Lévy process giving rise to the innovations. The parameters of the TML model is estimated by using a robust estimation method called the Expectation-Maximization (EM) algorithm.

A study of the Bole and Tamale historical DAT data showed that the deseasonalized DAT data has the mean-reversion property. Using plots and test statistics, it was observed that the residuals of the deseasonalized data are not normally distributed. To model the nonnormality in the residuals, we employed the generalized hyperbolic, normal-inverse Gaussian, and hyperbolic distributions to capture the skewness and semiheavy tails in the residuals. The hyperbolic distribution was found to be the best distribution that can capture the semiheavy tails and skewness in the empirical distributions of the residuals in the shifted ("extreme") regime. The introduction of the generalized hyperbolic and its subclasses led us to use the Lévy process in the shifted regime of the TML model for the deseasonalized temperature. The proposed regime-switching model is flexible as it modelled the deseasonalized temperature data reasonably well. Also, it was observed that there are instances that the temperature at both measurement stations are at their extremes as stated in the introduction.

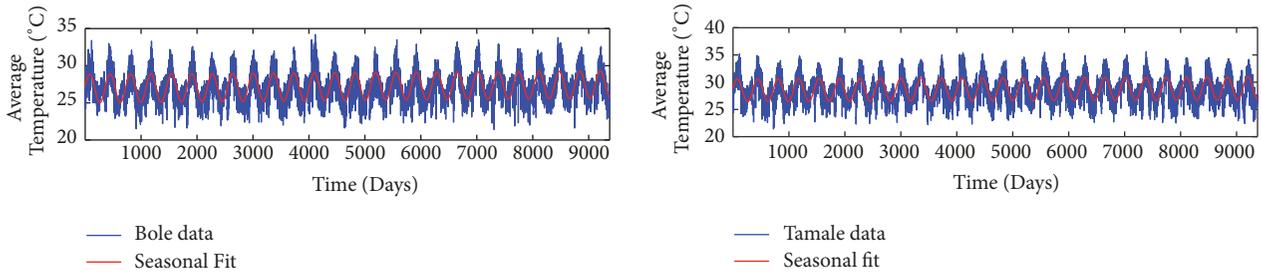

Figure 8: Seasonal fit plot of Bole and Tamale residuals.

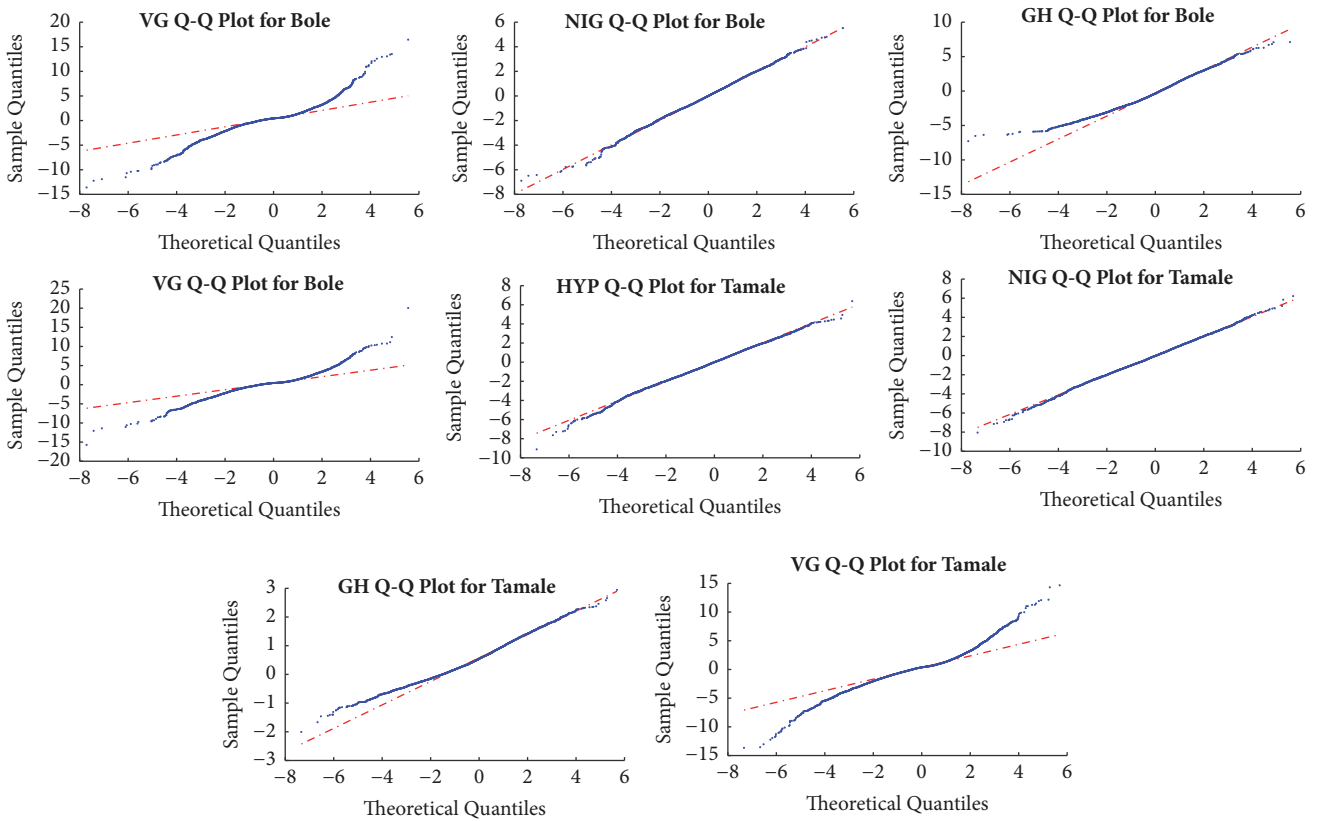

Figure 9: Q-Q plots of HYP, NIG, GH, and VG distributions.

From our results, it is evident that, due to the changes in volatility of the daily average temperature dynamics, conventional models that depend on Gaussian distribution will be ineffective when pricing weather derivatives in the derivative market. Our model however can be used to price weather derivative contracts written on temperature indices (CAT and GDD) for farmers in Africa since it incorporates the stylized facts of temperature in the model.

## Data Availability

The data used to support the findings of this study are available from the corresponding author upon request.

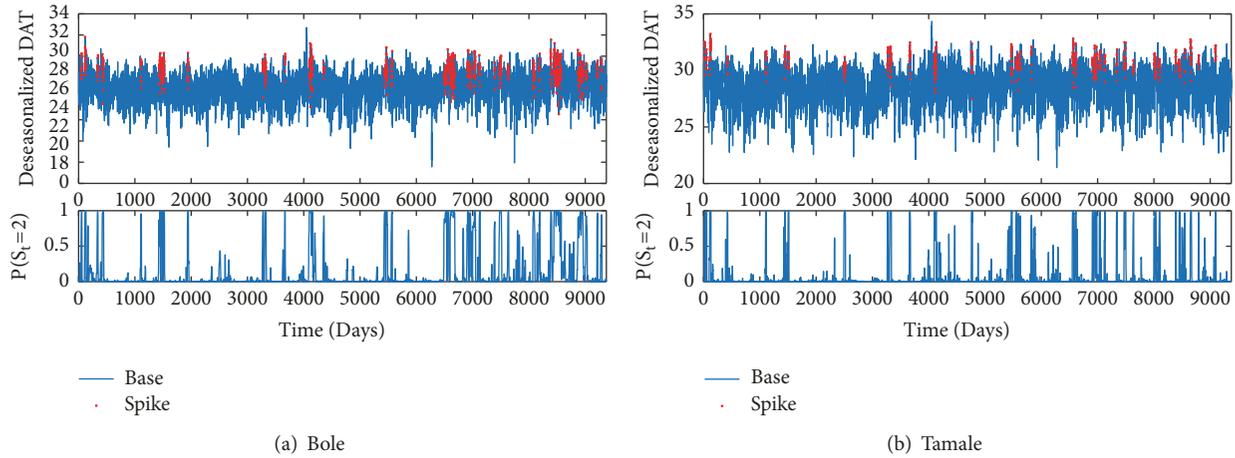

FIGURE 10: Calibration results of the MRS model with two independent regimes fitted to the deseasonalized daily average temperature. The lower panel shows the conditional probability $\mathbb{P}$ of being in the extreme regime.

TABLE 6: Estimated parameters for deterministic seasonality model.

|  | $A_0$ | $A_1$ | $A_2$ | $\varphi$ |
|---|---|---|---|---|
| Bole | 26.8194 | $2.3855 \times 10^{-5}$ | -2.0234 | 196.2153 |
| Tamale | 28.5058 | $3.7039 \times 10^{-05}$ | -2.1026 | 200.5695 |

TABLE 7: Estimated parameters of Normal, HYP, GH, NIG, and VG distributions fitted to Bole and Tamale DAT residuals. The parameters are estimated using the maximum likelihood method.

|  | Normal | HYP | GH | NIG | VG |
|---|---|---|---|---|---|
| **Bole** | | | | | |
| $\nu$ | – | – | 3.2875 | – | 0.0144 |
| $\alpha$ | – | 1.7178 | $1.4813 \times 10^{-5}$ | 1.5010 | 0.4968 |
| $\beta$ | – | -0.3921 | -0.1839 | -0.4087 | -0.0140 |
| $\mu$ | $1.0873 \times 10^{-6}$ | 0.6179 | 0.4638 | 0.6413 | 0.5004 |
| $\delta$ | 1.2995 | 1.6783 | $5.5849 \times 10^{-6}$ | 2.2664 | 0 |
| **Tamale** | | | | | |
| $\nu$ | – | – | 3.7329 | – | 0.0101 |
| $\alpha$ | – | 1.6520 | $3.1272 \times 10^{-5}$ | 1.5212 | 0.5006 |
| $\beta$ | – | -0.5406 | -0.2570 | -0.5807 | -0.0097 |
| $\mu$ | $3.6874 \times 10^{-6}$ | 1.1181 | 0.6802 | 1.1893 | 0.4793 |
| $\delta$ | 1.5431 | 2.2130 | $1.0713 \times 10^{-5}$ | 2.8794 | 0 |

TABLE 8: Estimated parameters for TML model.

| Parameter | $\sigma_1$ | $\kappa$ | $\mu$ | $\sigma_2$ | $p_{11}$ | $p_{22}$ |
|---|---|---|---|---|---|---|
| Bole | 0.0656 | 0.2047 | 29.8465 | 1.3939 | 0.9913 | 0.9490 |
| Tamale | 0.0100 | 0.3574 | 31.7835 | 1.4579 | 0.9909 | 0.9135 |

TABLE 9: Goodness-of-fit test using the Kolmogorov-Smirnov and Anderson-Darling.

|  | Normal | HYP | GH | NIG | VG |
|---|---|---|---|---|---|
| **Bole** | | | | | |
| Kolmogorov-Smirnov | 2.9650 | **0.6474** | 1.8743 | 0.7000 | 3.8201 |
| Anderson-Darling | 26.0323 | **0.5346** | 10.6454 | 0.7398 | 36.3417 |
| **Tamale** | | | | | |
| Kolmogorov-Smirnov | 1.4567 | **0.8977** | 1.2440 | 0.9370 | 2.4174 |
| Anderson-Darling | 36.0366 | **0.6035** | 8.8242 | 0.7864 | 41.6297 |

## Conflicts of Interest

## Acknowledgments

The authors wishes to thank African Union and Pan African University and Institute for Basic Sciences Technology and Innovation, Kenya, for their financial support for this research.

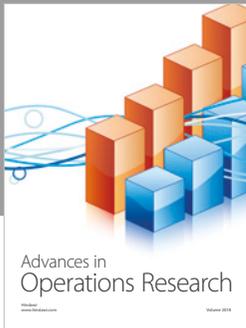 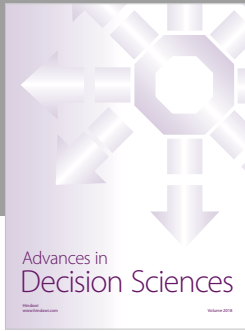 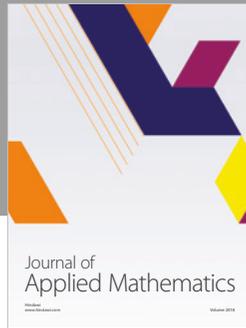 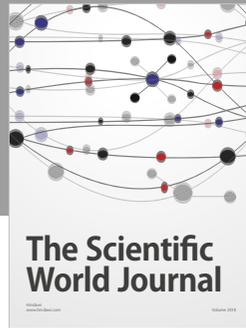 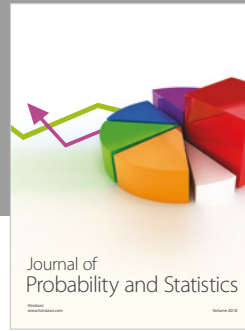
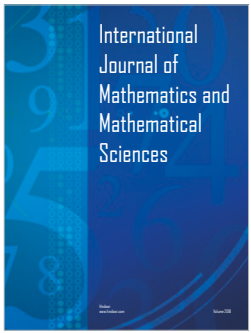 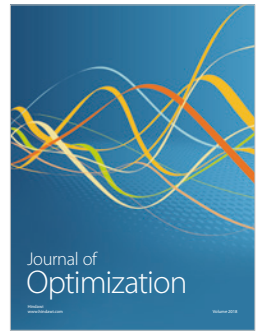
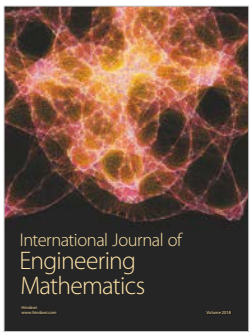 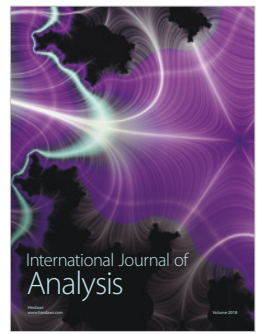
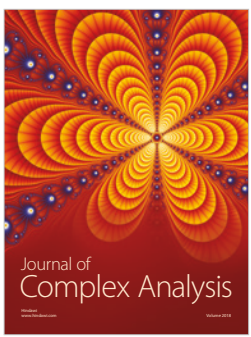 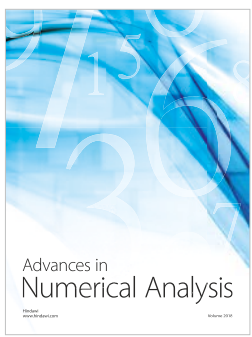 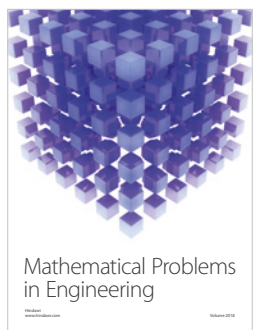 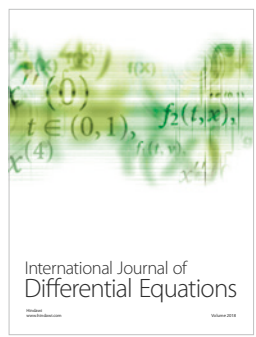 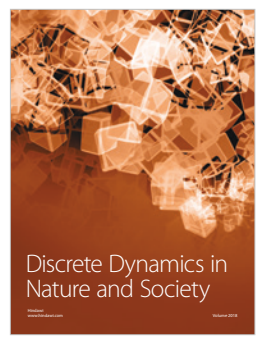
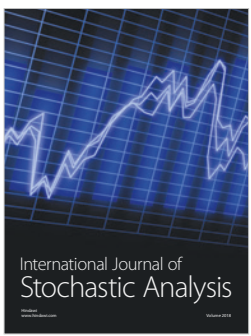 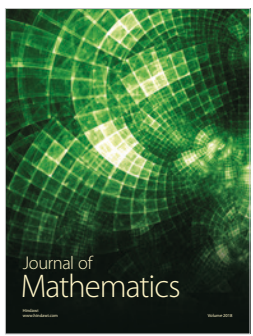 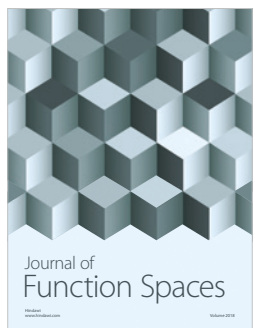 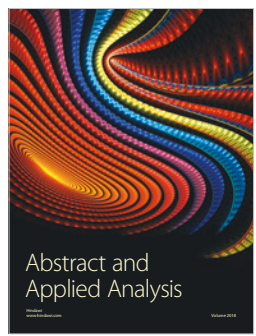 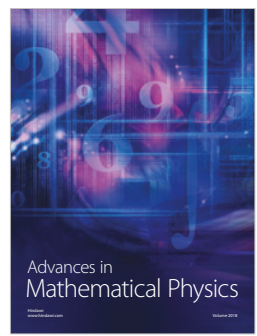

Submit your manuscripts at
www.hindawi.com